\documentclass[aps,amsfonts,nofootinbib,twocolumn,superscriptaddress]{revtex4-1}

\usepackage{amsmath}
\usepackage{amsbsy}
\usepackage{amssymb} 
\usepackage{color}
\usepackage{bbm}

\newcommand{\bb}{\begin{equation}}
\newcommand{\ee}{\end{equation}}

\def\sc{\scriptscriptstyle}

\begin{document}
\title{Generalized Dirac duality and $\mathbf{CP}$ violation in a two photon theory}
\author{Paola Arias}
\email{paola.arias@usach.cl}
\affiliation{Departamento de  F\'{\i}sica, Universidad de  Santiago de
  Chile, Casilla 307, Santiago, Chile}
 \author{Ashok K. Das}
 \email{das@pas.rochester.edu}
 \affiliation{ Department of Physics and Astronomy, University of Rochester,  Rochester, NY 14627-0171, USA}
\affiliation{ Institute of Physics, Sachivalaya Marg, Bhubaneswar 751005, India}
\author{Jorge Gamboa}
\email{ jgamboa55@gmail.com}
\affiliation{Departamento de  F\'{\i}sica, Universidad de  Santiago de
  Chile, Casilla 307, Santiago, Chile}
\author{Fernando M\'endez}
 \email{fernando.mendez@usach.cl}
\affiliation{Departamento de  F\'{\i}sica, Universidad de  Santiago de
  Chile, Casilla 307, Santiago, Chile}

\pacs{12.60.-i, 11.30.Cp}
\begin{abstract}
A  kinetic mixing term, which generalizes the duality symmetry of Dirac, is studied in a theory with two photons (visible and hidden). This theory can be either $CP$ conserving or $CP$ violating depending on the transformation of fields in the hidden sector. However if $CP$ is violated, it necessarily occurs in the hidden sector. This opens up an interesting possibility of new sources of $CP$ violation. 
\end{abstract}

\maketitle

It is known by now, through indirect evidence, that our universe contains a large amount of dark matter and 
the need to establish the existence of dark matter observationally is one of the most pressing issues at present\cite{review}. Since dark matter interacts very weakly with visible matter, any direct observation of processes involving the dark (hidden) sector needs to have the signals amplified enough so as to be accessible to measurements. However, after many interesting efforts in this direction it seems that new ideas are needed to detect the dark sector of our universe \cite{review2}. 

An alternative approach is to look for modifications induced in the observable phenomena in the visible sector due to the presence of a hidden sector in the physical theory \cite{andreas}. Of course, the main difficulty is that the structure of the theory describing the hidden sector is not very well understood at present. So, various extensions of the standard model have been proposed in recent years. One such model considers extending the standard $U(1)$ group of electromagnetism to $U(1) \times U_{h}(1)$ containing an additional hidden photon \cite{holdom}. The hidden photon mixes with the visible photon through a kinetic mixing term so that the Lagrangian density has the form \cite{holdom, p1}
\begin{equation}
{\cal L} = - \frac{1}{4}\,F_{\mu\nu}F^{\mu\nu} - \frac{1}{4}\,G_{\mu\nu}G^{\mu\nu} - \frac{\gamma}{2}\,F_{\mu\nu}G^{\mu\nu},\label{holdom}
\end{equation}
where the field strength tensors are defined as 
\bb 
F_{\mu \nu}=\partial_\mu A_\nu - \partial_\nu A_\mu,\quad G_{\mu \nu}=\partial_\mu X_\nu - \partial_\nu X_\mu. \label{a1}
\ee 
Here $A_\mu\in U(1)$ represents the visible photon field while $X_\mu\in U_{h}(1)$ corresponds to the photon field in the hidden sector and $\gamma$ is a small parameter denoting the strength of mixing between the two sectors. The Lagrangian density \eqref{holdom} has a \lq\lq mirror" symmetry between the visible and the hidden sectors, namely, it is invariant under 
\begin{equation}
A_{\mu}\leftrightarrow X_{\mu}.\label{mirrorsymmetry}
\end{equation} 
This model has been widely studied \cite{andreas,p1,wide} leading to several interesting features such as charge shift, millicharge particles and so on which result from the kinetic mixing of the photon with a hidden sector photon in \eqref{holdom}. 

In this letter we consider an alternative kinetic mixing term motivated by Dirac's duality symmetry \cite{dirac} in Maxwell's equations and study its consequences. Let us recall that the Maxwell's equations, if there were magnetic charges and currents, will have the form
\begin{equation}
\partial_{\mu}F^{\mu\nu} = J^{\nu},\quad \partial_{\mu}\widetilde{F}^{\mu\nu} = J^{\nu}_{m},\label{ME}
\end{equation}
where $\widetilde{F}^{\mu\nu} = \frac{1}{2} \epsilon^{\mu\nu\lambda\rho} F_{\lambda\rho}$ denotes the dual of the field strength tensor $F_{\mu\nu}$ and $J^{\mu}, J^{\mu}_{m}$ denote the electric and magnetic current densities respectively. Dirac had observed that in the presence of magnetic currents, Maxwell's equations \eqref{ME} would be invariant under the duality symmetry
\begin{equation}
F_{\mu\nu}\leftrightarrow \widetilde{F}_{\mu\nu},\quad J^{\mu}\leftrightarrow J^{\mu}_{m},\label{duality}
\end{equation}
leading to a quantization of electric charge  \cite{dirac,schwinger,goddard,preskill}. Note that the duality symmetry \eqref{duality} manifests only at the level of equations of motion \eqref{ME} and not in the Lagrangian density for the Maxwell theory. Of course, monopoles have not been observed yet \cite{kimball} and, therefore, this discussion of a duality symmetry may  seem rather academic.
 
On the other hand, in an enlarged theory with the symmetry group $U(1)\times U_{h}(1)$, Dirac's duality symmetry can be naturally generalized in the following way. Consider the Lagrangian density describing the two photons to be
\bb 
{\cal L} = -\frac{1}{4} F_{\mu \nu}  F^{\mu \nu}  -\frac{1}{4} G_{\mu \nu}  G^{\mu \nu} - \frac{\gamma}{2}\,F_{\mu\nu}\widetilde{G}^{\mu\nu},\label{a2} 
\ee
where the field strength tensors $F_{\mu \nu}$ and $G_{\mu \nu}$ are defined in \eqref{a1} and $\widetilde{G}^ {\mu \nu} =\frac{1}{2} \epsilon^{\mu \nu \lambda \rho} G_{\lambda \rho}$ denotes the dual of $G_{\mu\nu}$. As before, we assume that the strength $\gamma$ of the kinetic mixing term is small. Therefore, in contrast to the conventional kinetic mixing in \eqref{holdom}, here we have a coupling involving the dual field strength tensor. (We note here that such a mixing term has been considered before from a different perspective \cite{jae} where it is called a \lq\lq magnetic mixing" term.) It is useful to point out that the kinetic mixing term in \eqref{a2} can also be written as $\widetilde{F}_{\mu\nu}G^{\mu\nu}$. (The mixing term in \eqref{a2} is reminiscent of an axion coupling term to photons \cite{pq}.) At first sight, this appears to be the same Lagrangian density as in \eqref{holdom} because under the redefinition 
\begin{equation}
F_{\mu\nu}\rightarrow -\widetilde{F}_{\mu\nu},\quad \text{or},\quad G_{\mu\nu}\rightarrow -\widetilde{G}_{\mu\nu},\label{redef1}
\end{equation}
the kinetic mixing term in \eqref{a2} goes over to the one in \eqref{holdom}. However, under \eqref{redef1}, one of the kinetic energy terms in \eqref{a2} changes sign signalling an unbounded energy (for example, $\widetilde{G}_{\mu\nu}\widetilde{G}^{\mu\nu} = - G_{\mu\nu}G^{\mu\nu}$). So, in fact, the kinetic mixing term in \eqref{a2} is different from the one in the Holdom model \eqref{holdom}.

Since there is no monopole in the visible world, for consistency with observations, we choose
\begin{equation}
\partial_{\mu} \widetilde{F}^{\mu\nu} = 0.\label{nomonopole}
\end{equation}
On the other hand, we do not know anything about the hidden sector and a monopole may, in fact, exist in this  (hidden) sector which allows us to choose
\begin{equation}
\partial_{\mu}\widetilde{G}^{\mu\nu} \equiv J^{\nu} \neq 0.\label{Gmonopole}
\end{equation}
In fact, we need a condition like \eqref{Gmonopole} for the kinetic mixing term to be nontrivial. (The mixing term in \eqref{a2} will be a total divergence with \eqref{nomonopole} if $\widetilde{G}^{\mu\nu}$ were also divergenceless and will not lead to any change in dynamics.) We note here that, while the Lagrangian density \eqref{a2} is invariant under the mirror symmetry transformation \eqref{mirrorsymmetry}, equations \eqref{nomonopole} and \eqref{Gmonopole} violate this  symmetry.

On the other hand, dynamical equations for this coupled set of photons develop a generalized duality symmetry. To see this, let us note that, together with \eqref{nomonopole} and \eqref{Gmonopole}, the dynamical equations following from \eqref{a2} are given by
\begin{align} 
\partial_\mu F^{\mu \nu} & = -\gamma J^\nu,\quad & \partial_{\mu}\widetilde{F}^{\mu\nu} & = 0,\label{a3}\\
\partial_\mu G^{\mu \nu} & = 0,\quad &\partial_{\mu}\widetilde{G}^{\mu\nu} & = J^{\nu}, \label{a4}
\end{align}
where, in \eqref{a3}, $J^\nu = \partial_\mu \widetilde{G}^{\mu \nu}$ (see \eqref{Gmonopole} or \eqref{a4}). 

Basically  the first equation in  \eqref{a3} is the usual Maxwell equation coupled to a source produced by the hidden photons resulting from the kinetic mixing term in \eqref{a2}. The second equation in \eqref{a3} is the no monopole condition of \eqref{nomonopole}. The first equation in \eqref{a4} corresponds to the Maxwell equation for the hidden photon without any source since the kinetic mixing term does not contribute to a source for the dark sector with the assumption in \eqref{nomonopole}. The second equation in \eqref{a4} represents our assumption \eqref{Gmonopole}.  

The set of dynamical equations \eqref{a3}-\eqref{a4} are not invariant under the \lq\lq mirror" symmetry \eqref{mirrorsymmetry} even though the Lagrangian density \eqref{a2} is. This can be traced back to our assumption of the existence of monopoles in the hidden sector (namely, the violation of Bianchi identity in \eqref{Gmonopole}). Namely, even though the Lagrangian density \eqref{a2} is \lq\lq mirror" symmetric, the conditions on the dual field strength tensors in \eqref{nomonopole} and \eqref{Gmonopole} violate this symmetry explicitly. 

On the other hand, the dynamical equations \eqref{a3}-\eqref{a4} develop a generalized duality symmetry. Namely, they are invariant under 
\begin{align} 
 F^{\mu \nu} & \leftrightarrow -\gamma\, \widetilde{G}^{\mu \nu}, \nonumber\\ 
 J^\mu & \leftrightarrow J^\mu. \label{a5}
\end{align}
Thus, we see that the kinetic mixing term in \eqref{a2}, together with \eqref{nomonopole}-\eqref{Gmonopole}, generalizes Dirac's duality symmetry to this two photon theory. Here the duality is between photons in {\it the visible and the hidden sectors} which is different from Dirac's original duality.  
 
As we have mentioned earlier, the kinetic mixing term in \eqref{a2} is reminiscent of the axion coupling involving the dual field strength tensor. Therefore, one should analyze the behavior of the theory under a $CP$ transformation. It is well known that the visible electric and magnetic fields  transform under $CP$ as
\begin{equation}
\mathbf{E} \xrightarrow{\sc\mathbf{CP}} \mathbf{E},\quad \mathbf{B} \xrightarrow{\sc\mathbf{CP}} -\mathbf {B}, \label{visibleCP}
\end{equation}
and the kinetic energy term $F_{\mu\nu}F^{\mu\nu}$ in \eqref{a2} is invariant under $CP$. However, we do not know how the electric and magnetic fields in the hidden sector transform under $CP$, except that each of $\mathbf{E}$ and $\mathbf{B}$ should transform either as a vector or as a pseudovector. Thus, we can assume the following general $CP$ transformations for the electric and the magnetic fields in the hidden sector 
\begin{equation} 
\mathbf{E}_h \xrightarrow{\sc\mathbf{CP}} \,\alpha\, \mathbf{E}_h,\quad \mathbf{B}_h \xrightarrow{\sc\mathbf{CP}} \,\beta\, \mathbf{B}_h, \label{hiddenCP}
\end{equation}
where $\alpha,\beta=\pm 1$. The kinetic energy term  $G_{\mu \nu}G^{\mu \nu}$ in \eqref{a2} is $CP$ invariant, as it should be,  with these values of the parameters $\alpha,\beta$  
\[
(\mathbf{E}_{h}^{2} - \mathbf{B}_{h}^{2})\xrightarrow{\sc\mathbf{CP}} (\alpha^2 \mathbf{E}_h^2 -\beta^2  \mathbf{B}_h^2) = (\mathbf{E}_h^2 - \mathbf{B}_h^2). 
\] 

Of the four possible choices $\alpha=1, \beta=\pm1$ and $\alpha=-1, \beta=\pm 1$, only two are compatible with Lorentz invariance. Namely, we can only have either 
\begin{equation}
\alpha = 1, \beta = -1,\quad \text{or,}\quad \alpha = -1, \beta = 1,\label{parameters}
\end{equation}
for the transformations in \eqref{hiddenCP}. With $\alpha=1, \beta=-1$ the electric and magnetic fields in the hidden sector will behave exactly like the ones in the visible sector (see \eqref{visibleCP}) while with $\alpha=-1,\beta=1$, the hidden electric and magnetic fields would have the opposite behavior from the visible sector under $CP$. The first choice in \eqref{parameters} would be consistent with the \lq\lq mirror" symmetry \eqref{mirrorsymmetry} while the second choice is more in line with the generalized duality symmetry \eqref{a5}.

Both the kinetic energy terms in \eqref{a2} are invariant with either of the choices in \eqref{parameters}. However, the kinetic mixing term $F_{\mu\nu}\widetilde{G}^{\mu\nu}$ behaves in an opposite maner depending on the choice. We note that under a $CP$ transformation \eqref{visibleCP} and \eqref{hiddenCP}
\begin{equation}
(\mathbf{E}\cdot \mathbf{B}_{h} + \mathbf{E}_{h}\cdot \mathbf{B})\xrightarrow{\sc\mathbf{CP}} (\beta \mathbf{E}\cdot \mathbf{B}_{h} - \alpha \mathbf{E}_{h}\cdot \mathbf{B}),\label{mixing}
\end{equation} 
so that with $\alpha =1$ and $\beta =-1$  
\begin{equation} 
(\mathbf{E}\cdot \mathbf{B}_{h} + \mathbf{E}_{h}\cdot \mathbf{B})\xrightarrow{\sc\mathbf{CP}} - (\mathbf{E}\cdot \mathbf{B}_{h} + \mathbf{E}_{h}\cdot \mathbf{B}),
\end{equation} 
and the kinetic mixing term (and, therefore, the total Lagrangian density) violates $CP$. On the other hand, with $\alpha = -1, \beta=1$ we have
\begin{equation}
(\mathbf{E}\cdot \mathbf{B}_{h} + \mathbf{E}_{h}\cdot \mathbf{B})\xrightarrow{\sc\mathbf{CP}} (\mathbf{E}\cdot \mathbf{B}_{h} + \mathbf{E}_{h}\cdot \mathbf{B}),
\end{equation} 
and the kinetic mixing term (and, therefore, the total Lagrangian density) is $CP$ invariant. This shows that the theory \eqref{a2} can be either $CP$ conserving or $CP$ violating depending on the two possible choices for the transformation of the hidden electric/magnetic fields. However, in either case, the theory is $CPT$ invariant.

This can be contrasted with the model in \eqref{holdom} where the manifest \lq\lq mirror" symmetry naturally picks the first choice in \eqref{parameters} and a mixing term of the form $F_{\mu\nu}G^{\mu\nu}$ leads to a $CP$ conserving theory, namely, for $\alpha=1,\beta=-1$,
\begin{align}
(\mathbf{E}\cdot \mathbf{E}_{h} + \mathbf{B}\cdot \mathbf{B}_{h}) & \xrightarrow{\sc\mathbf{CP}} (\alpha\mathbf{E}\cdot\mathbf{E}_{h} - \beta\mathbf{B}\cdot \mathbf{B}_{h}) \nonumber 
\\
&\ =\ (\mathbf{E}\cdot \mathbf{E}_{h} + \mathbf{B}\cdot \mathbf{B}_{h}). \nonumber
\end{align}

The $CP$ violation in a theory of photons may seem problematic because of the existence of stringent experimental bounds. However, if the violation occurs only in the hidden sector, it should, in principle, be fine. We note that the Lagrangian density \eqref{a2} can be diagonalized with the redefinition of the hidden photon field strength as
\begin{equation}
\overline{G}_{\mu\nu} = G_{\mu\nu} + \gamma \widetilde{F}_{\mu\nu},\label{redef}
\end{equation}
so that it has the form
\begin{equation}
{\cal L} = - \frac{1}{4} (1+\gamma^{2}) F_{\mu\nu}F^{\mu\nu} - \frac{1}{4} \overline{G}_{\mu\nu} \overline{G}^{\mu\nu}.\label{diagonalized}
\end{equation}
This redefinition decouples the visible and the hidden sector with a rescaling of the visible electric charge (this rescaling is similar to that obtained for \eqref{holdom} upon diagonalization except that, in that case, the photon kinetic energy is rescaled by $(1-\gamma^{2})$). However, more importantly, since $F_{\mu\nu}F^{\mu\nu}$ is $CP$ conserving, there is no $CP$ violation in the visible sector of the decoupled theory. All the $CP$ violation, resulting from the first choice of parameters in \eqref{parameters}, has been shifted into the hidden sector. This, in fact, opens up a new possible source of $CP$ violation.
\medskip

This work was supported by FONDECYT/Chile grants and 1161150 (P.A.), 1130020 (J.G.), 1140243 (F.M.) and Dicyt-USACH (J.G.). We would like to thanks  Professors J. L.  Cort\'es and   H. Falomir  for discussions.

\end{document}